\documentclass[aps,prl,preprint]{revtex4-1}
\usepackage{color}
\usepackage{colortbl,amsthm,amsmath,amssymb,txfonts}
\usepackage{graphicx}
\usepackage{epsfig}
\usepackage{float}
\usepackage{siunitx}
\usepackage{textcomp}
\usepackage{gensymb}
\usepackage{subfigure}
\usepackage{caption}
\usepackage{multirow}

\begin{document}

 
\title{Precursors of the El Ni\~{n}o Phenomenon: A climate network analysis}
\author{Rupali Sonone$^1$}
\email{vaidehisonone@gmail.com}
\author{Neelima Gupte$^1$}
\email{gupte@physics.iitm.ac.in}
\address{$^{1}$Department of Physics, Indian Institute of Technology Madras, \\ Chennai, 600036, India.}

\begin{abstract}

The identification of precursors of climatic phenomena has enormous practical importance. Recent work constructs a climate network based on surface air temperature data to analyze the El Ni\~{n}o phenomena. We utilize  microtransitions which occur before the discontinuous percolation transition in the network as well as other network quantities  to identify a set of reliable precursors of El Ni\~{n}o episodes. These precursors identify nine out of twelve El Ni\~{n}o episodes occurring in the period of $1979$ to $2018$ with a lead time varying from six to ten months. We also find indicators of tipping events in the data.
\end{abstract}

\date{\today}
\maketitle
The climate system is highly complex nonlinear dynamical system consisting of various subsystems such as the atmosphere, the ocean, the cryosphere, the biosphere, and the lithosphere. The nonlinear interactions within these subsystems lead to various climatic phenomena on multiple timescales such as cyclones, the monsoon and oceanic currents\cite{cn2}. Climatic phenomena have major consequences for ecological and economic events. Hence the prediction of climatic phenomena is of enormous practical utility, but is quite difficult due to the complex and nonlinear nature of the phenomena under consideration. Thus, the construction of reliable predictors of such events is useful and important.   
Recent methods proposed to analyze such complex climate systems include the construction of a complex network from climate data \cite{cn1,cn2,tsr}. These networks have been used  to forecast some important climate phenomena, such as the monsoon \cite{monsoon1,monsoon2}, the North Atlantic Oscillation and El Ni\~{n}o events \cite{ygh}-\cite{prl5}.\par 
These climate networks are constructed using  massive datasets consisting of the time series of weather observations at different geographic locations. Generally, the nodes are chosen to be geographic locations, and links are added based on the similarities between  the dynamics at pairs of nodes.  Recently, the utility of designing generic tools that reliably track the structural changes in the network has been highlighted. Tracking these structural changes in the climate network of temperatures reveals severe response to climatic events such as El Ni\~{n}o events. \par

  The El Ni\~{n}o Southern Oscillation (ENSO) is the most influential climate phenomenon on interannual time scales. It is marked by irregular warm (El Ni\~{n}o) and cold (La Ni\~{n}a) anomalies of sea surface temperatures (SST) from the long-term mean state. Events are defined as the consecutive overlapping of three-month periods of SST anomalies ($\geq +0.5 \degree C$) for warm (El Ni\~{n}o) events and ($\leq -0.5 \degree C$) for cold (La Ni\~{n}a) events in the Ni\~{n}o $3.4$ region (i.e., $5 \degree N$-$5 \degree S$, $120 \degree$-$170 \degree W$). The threshold is further broken down into Weak ($0.5 \degree C$ - $0.9 \degree C$ SST anomaly), Moderate ($1.0 \degree C$ - $1.4 \degree C$), Strong ($1.5  \degree C$ - $1.9 \degree C$) and Very Strong ($\geq 2.0 \degree C$) events. This phenomenon triggers many disruptions around the globe causing disastrous flooding in countries like Peru and Ecuador as well as heavy droughts in large areas of South America, Indonesia, and Australia. It is plausibly also associated with severe winters in Europe, droughts in India, China, and Brazil, intense tropical cyclones in the Caribbean, and epidemic diseases occurring in various places \cite{nino,ninobook}. The ENSO phenomenon is currently tracked and quantified by the NINO $3.4$ index. Here, we use various datasets of the daily near surface ($1000$hPa) air temperature of ERA-Interim reanalysis (The European Centre for Medium-Range Weather Forecasts (ECMWF)) within the period $1979$-$2018$ to forecast the development of the El Ni\~{n}o  well in advance \cite{ecmwf}. Such an early forecast of the  El Ni\~{n}o event can contribute to  better disaster management planning and efficient distribution of resources well before the calamity.\par
   
Complex network systems, ranging from ecosystems to financial markets and the climate, can have tipping points at which a phase transition to a contrasting dynamical regime occurs. A variety of indicators, such as the order parameter, and the susceptibility, quantify the loss of resilience occurring generically when dynamical network systems approach transition point \cite{rep,rev}. The framework of percolation theory has been  used to analyze the behavior of connected clusters in the network and predict the transition point \cite{per}. Discontinuous percolation transitions have been observed in the climate network. We analyze these discontinuous percolation transitions and try to identify a set of  early warning precursors to these transitions, and use them to predict the onset of El Ni\~{n}o activity. We note that our precursors are able to predict El Ni\~{n}o activity  from six to ten months in advance. Our methods are quite reliable, and only lead  to one false positive prediction in the analysis of 39 years of data. Our methods consist of the analysis of microtransitions that occur before the discontinuous percolation transition and the topological analysis of network quantities such as node degrees, link densities and link lengths. 
\par

  The existence and predictive utility of microtransitions has been the focus of recent work in the context of percolation phenomena. These microtransitions appear before the percolation transition for both the continuous and discontinuous cases and accumulate at the transition point. These microtransitions are signalled by small jumps in the order parameter, i.e. the size of the maximal cluster,  and peaks in the variance of the order parameter. In the case of the climate network, the actual transition point is the point at which a discontinuous percolation transition is seen, with the usual jump in the order parameter. Here, we use the  microtransitions to predict critical thresholds at which the percolation  transition can occur \cite{micro}. The predicted threshold values are characteristic of the data of the period for which the network is constructed and can hence be used as  generic early-warning indicators of the El Ni\~{n}o years. The variance in the order parameter, i.e. the susceptibility at the critical point can also be used to identify pre-El Ni\~{n}o years (which we will call indicator years) , and thereby predict the El Ni\~{n}o in the subsequent year. We note that the jump in the order parameter has been used earlier to identify pre-El Ni\~{n}o years \cite{1}, but the susceptibility constitutes a stronger indicator. Additionally, we have carried out the topological analysis of the climate network of each year and have  found that several  distinct features such as the distribution of network links, the degrees of the network nodes in the El Ni\~{n}o basin   differentiate clearly  between El Ni\~{n}o years and indicator years. Thus, this entire collection of  features i.e. the susceptibility, critical correlation value, maximum value of correlation strength and total number of links observed in the climate network can be used as a set of precursors of an upcoming El Ni\~{n}o event. \\ \par

The climate network is constructed by the following method. We use daily near surface ($1000$hPa) air temperature datasets [$\tilde{T}^{y}(d)$] recorded at $1,15,680$  nodes whose geographical  location is fixed by a pair of latitude and longitude values. The grid size of the network can be changed by altering the resolution of nodes such that every element in the grid covers the same area on the globe. Thus the total number of nodes on the  equator is given by, $n_{0}= 360^{\degree}/r_{0}$, where $r_0$ is the grid resolution, and the total number of nodes on $mr_0$ latitude is given by $n_{m}=n_{0} \cos (mr_0) $ where $ m \in [- 90/r_0, 90/r_0]$. The total number of nodes in the climate network is then $N = \sum_{m=0}^{m=90/r_0} 2n_{m}-n_{0}$. A geographical grid of $726$ nodes with $7.5 \degree$ grid resolution is constructed and links are added between the two nodes by calculating the Pearson correlation function between their temperatures. The filtered daily near surface air temperature $T^{y}(d)$ and the cross-correlation function $C^{y}_{i,j}$  are defined as follows \cite{1},
\begin{eqnarray}
T^{y}(d)&=&\frac{\tilde{T}^{y}(d)-mean(\tilde{T}(d))}{std(\tilde{T}(d))} \\
C^{y}_{i,j}(\tau)&=& \frac{\langle T^{y}_i(d-\tau)T^{y}_j(d) \rangle -\langle T^{y}_i(d-\tau)\rangle\langle T^{y}_j(d)\rangle}{\sqrt{(T^{y}_i(d-\tau)-\langle T^{y}_i(d-\tau)\rangle)^2}.\sqrt{(T^{y}_j(d)-\langle T^{y}_j(d)\rangle)^2}}
\end{eqnarray}
Temperature datasets over $39$ years from 1979 to 2018  have  been used to built a network for each year from January $1^{st}$ till $31^{st}$ December, and $\tau$ is the time lag between $0$ to $200$ days. For each year, the data for the previous year is also required on account of the time lag. The appropriate time lag and its impact is discussed in \cite{tau}. Here "$mean$" and "$std$" are the mean and standard deviation of the temperature on day `$d$' over all years. Only the temperature data points prior to  day `$d$' are considered. The averages ($\langle ~~ \rangle_{d}$) considered here are taken over $365$ days. The weight of the link between $i$ and $j$ is defined as the maximum of the cross-correlation function $max$($C^{y}_{i,j}(\tau)$). Here, links were added one by one with decreasing link strength, i.e. the links or edges with the highest weight (maximum of the cross-correlation function $C^{y}_{i,j}$), are added first to the network. Further  edges are added in order by decreasing weight.
Thus the network evolves as a function of $C$, as links of weights ranging from a  given value of $C$  upto the maximum  value of $C$ are added at each stage. Quantities such as the  order parameter, which is the normalized size of the largest cluster ($s_1$) and the susceptibility ($\chi$) for each distinct value of $C$ are studied. These are defined by the relations \cite{1}, \\
\begin{eqnarray}
s_1 &=& \frac{S_1}{N} \\
\chi &=& \frac{\sum_{s}^{'}s^{2} n_s(C)}{\sum_{s}^{'}s n_s(C)}
\end{eqnarray}
Here, $S_1$ is the size of largest cluster and $N$ is the total number of nodes, $n_s(C)$ is the number of clusters of size "$s$" for link’s weight $C$ and below, and the prime on the sums indicates the exclusion of the largest cluster $S_1$ in each measurement. The transition to the percolating state in this climate network is identified via  the existence of a giant cluster containing $O(N)$ nodes. As mentioned earlier, this quantity is identified as the order parameter\par

We note  that  El Ni\~{n}o events have been classified as very strong, strong and moderate events based on the El Ni\~{n}o index as explained above,  and compare the networks corresponding to each class. These networks are characterized by plotting the susceptibility ($\chi$) and the largest normalized cluster ($s_1$). A discontinuous phase transition is clearly observed in $s_1$ (the largest normalized cluster. We plot these quantities as functions of $C$ for typical  years with  El Ni\~{n}o events of each class, as well as the preceding years, which turn out to be indicator years of the event. \par
The year is indicated at the top of all the plots, and the red line marks the critical value $C_c$ on the $x$-axis at which the largest jump is observed in $s_1$. The transition to percolation is signaled by the value of $s_1$ tending toward one, and the existence of a maximal cluster. It can be clearly seen from the plots that $\Delta s_1$, the jump in the size of the maximal cluster, is largest for the indicator year i. e. one year prior to the El Ni\~{n}o episode. The susceptibility $\chi$ is also plotted on the same graphs. It can be seen that the magnitude of the jump $\Delta \chi $ is significantly larger in the indicator years, i.e. one year prior to the El Ni\~{n}o episodes compared to other years. 
\begin{figure}[H]
\begin{tabular}{c c}
\includegraphics[width=0.46\textwidth]{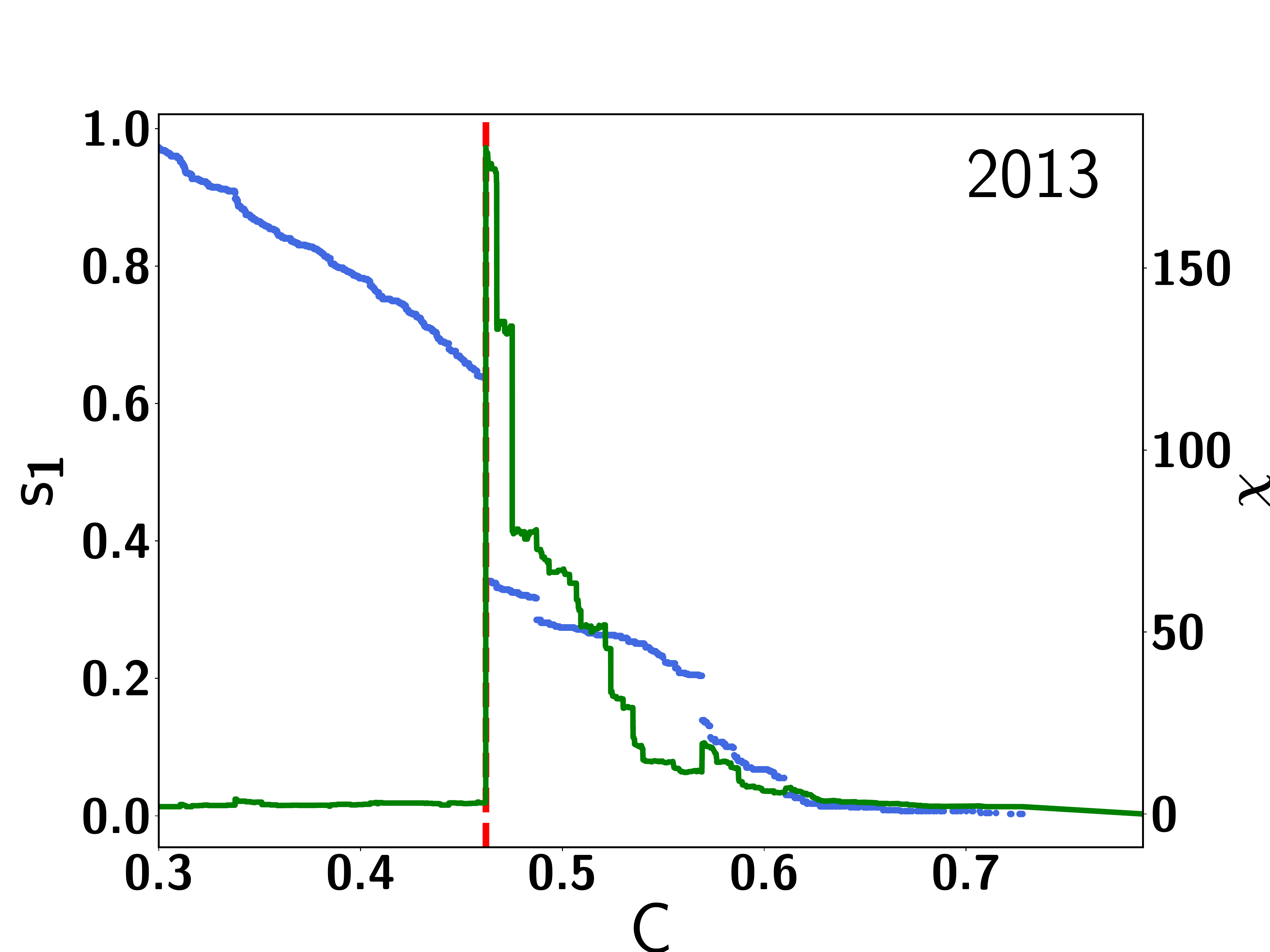}  &
\includegraphics[width=0.46\textwidth]{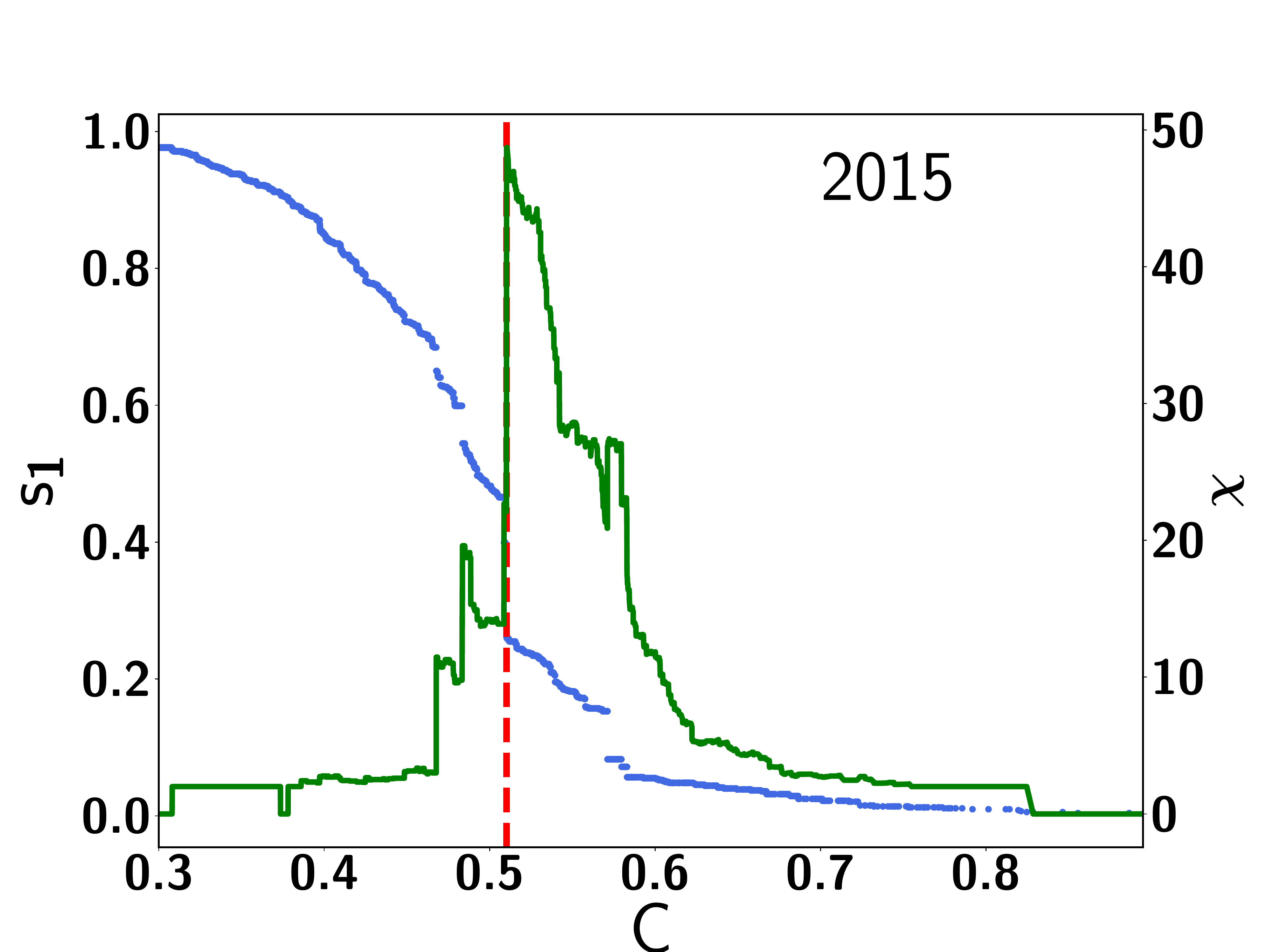}   \\
(a) & (d) \\
\includegraphics[width=0.46\textwidth]{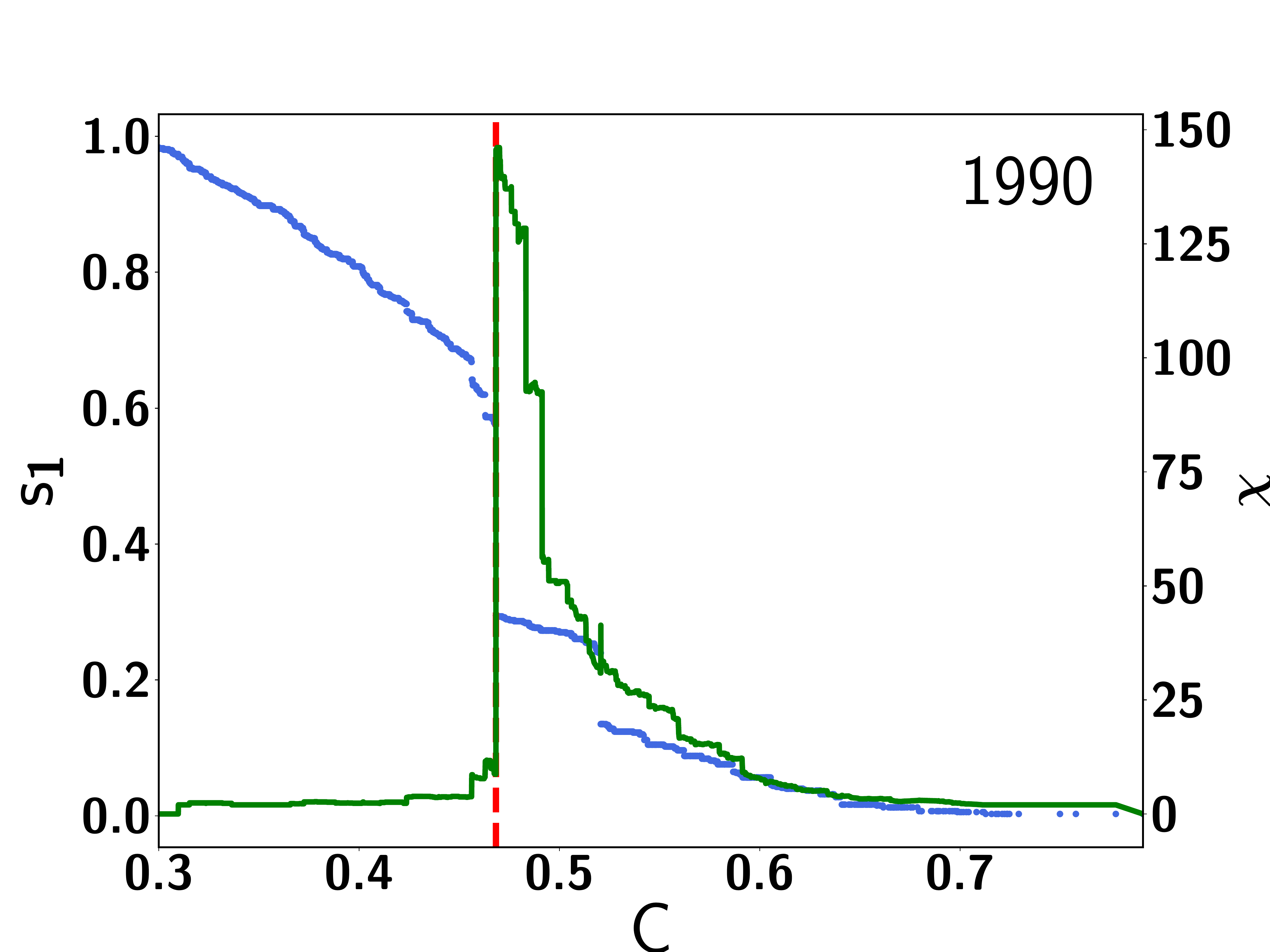}  &
\includegraphics[width=0.46\textwidth]{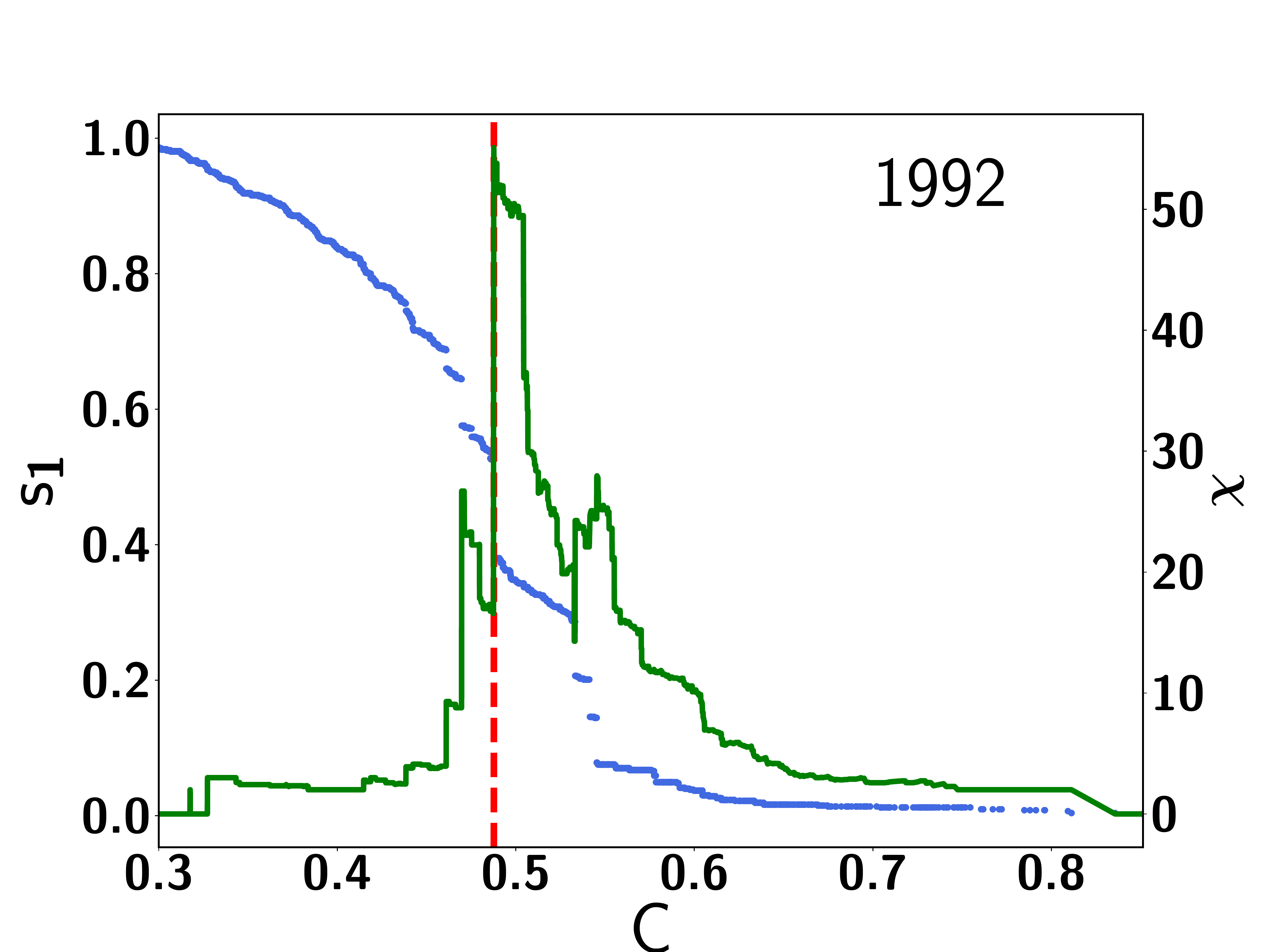}   \\
(b) & (e) \\
\includegraphics[width=0.46\textwidth]{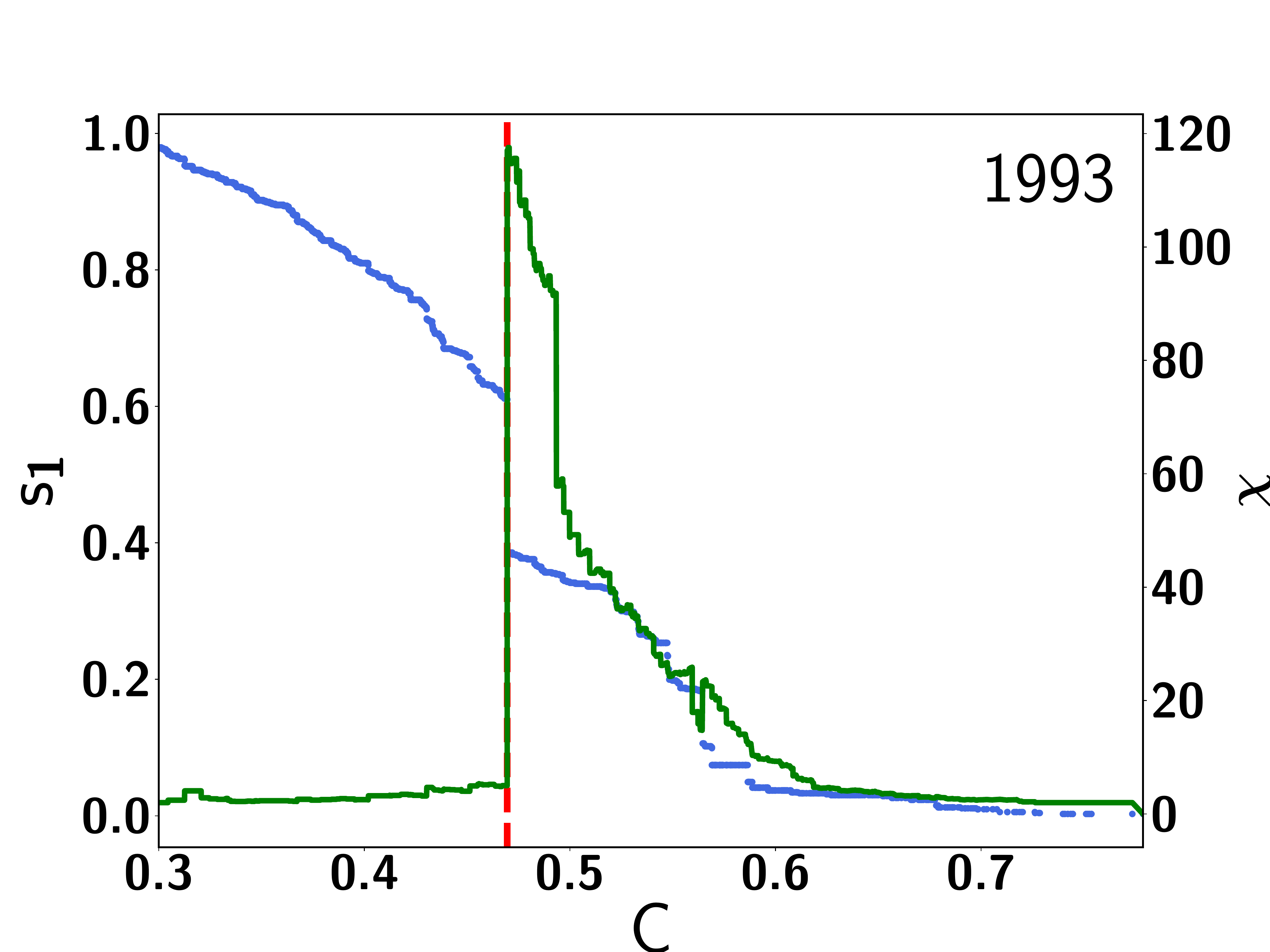}  &
\includegraphics[width=0.46\textwidth]{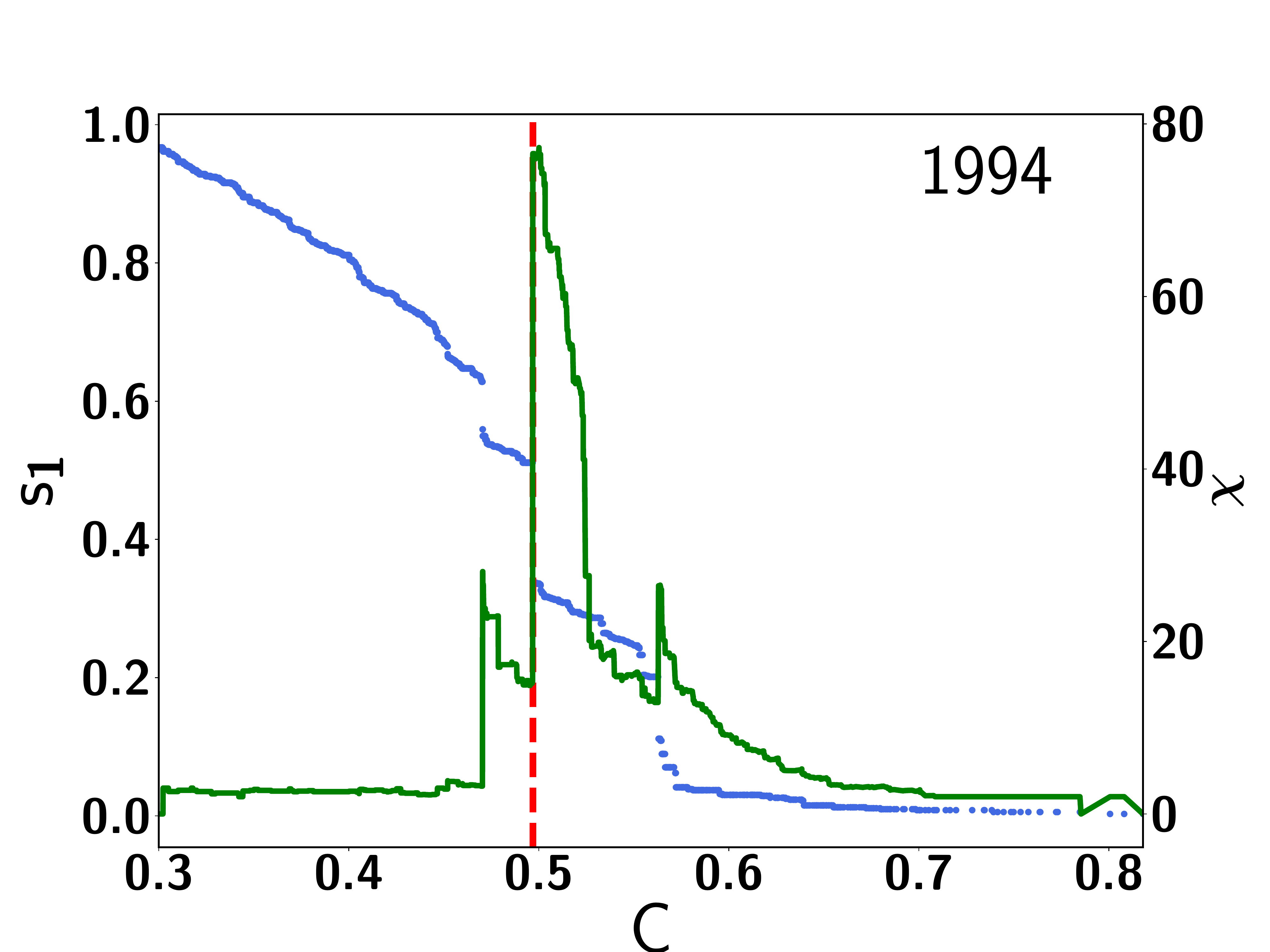}   \\
(c) & (f) \\

\end{tabular}
\caption{The largest cluster $s_1$ (blue dotted line) and susceptibility $\chi$ (green line) is plotted  as a function of the link strength $C$ for the network. Plots (a)(December 2013), (b)(December 1990) (c ) December 1993, which  are  years which precede  a very strong El Ni\~{n}o episode, strong El Ni\~{n}o episode, and a  moderate El Ni\~{n}o episode, respectively. Plots (d) (e) (f) are for years when the El Ni\~{n}o episode peaks, (d) very strong El Ni\~{n}o episode, December 2015. (e) strong El Ni\~{n}o episode, December 1992. (f) moderate El Ni\~{n}o episode, December 1994.}
\end{figure} 

 The susceptibility of the network gives the measure of the change in the formation of  clusters of varied sizes at a given  correlation strength. We note that while the order parameter jump has been suggested as a measure of the precursor \cite{1}, the susceptibility is a more definitive and reliable precursor for the upcoming El Ni\~{n}o episode as not only is the largest peak $\Delta \chi $ significant, but the peaks in the susceptibility before it also have a very distinct pattern. Ref.\cite{1} suggested that susceptibility has seemingly no distinct relation with the El Ni\~{n}o Index and hence refrained from using the susceptibility as a predictor. In contrast to their observations, it can be seen from the plots above that the susceptibility has a distinct pattern and clearly shows a series of small transitions i.e. microtransitions before the percolation transition point.
\begin{figure}[H]\label{Fig4}
\includegraphics[width = 12.0cm, height = 7cm]{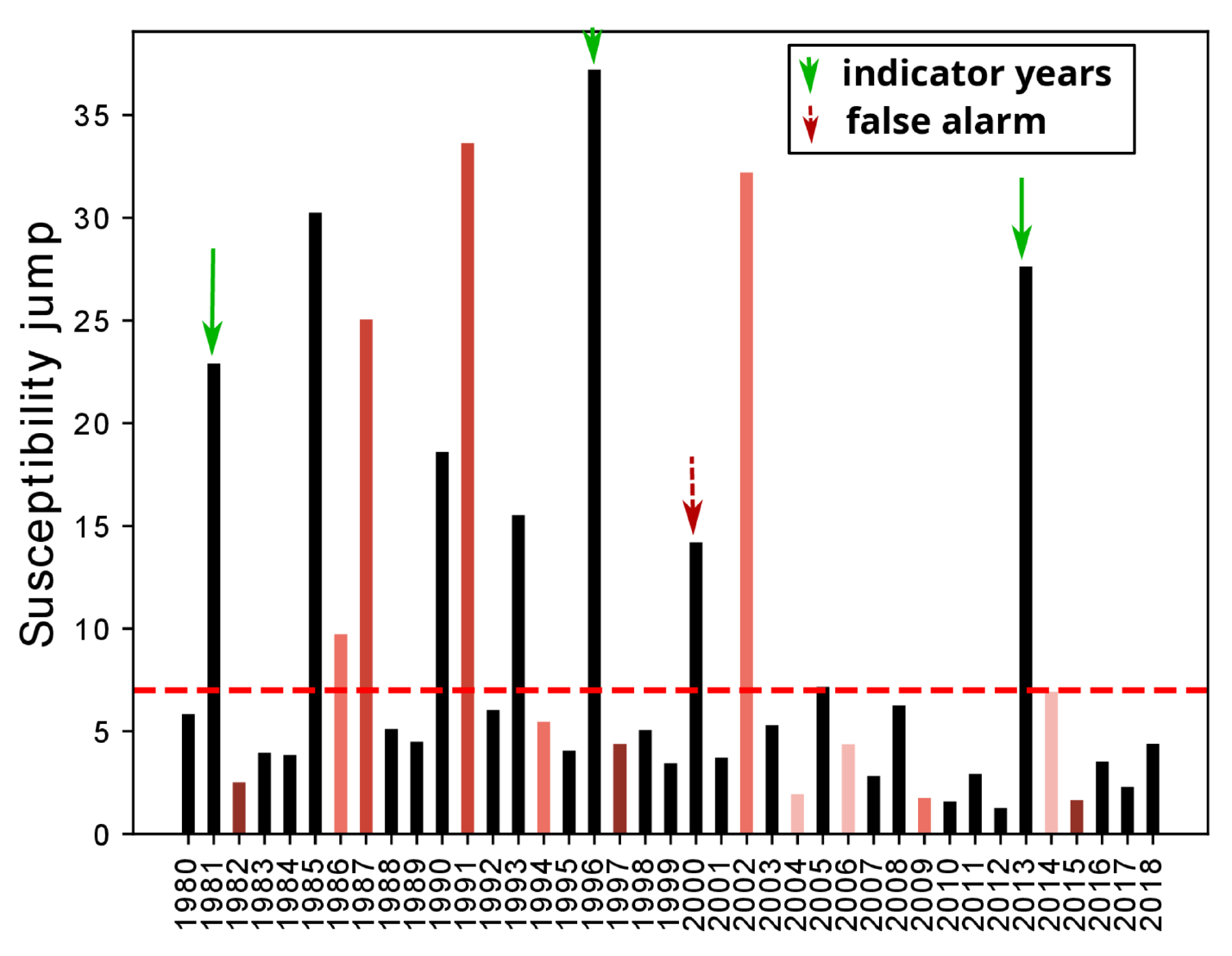} \\
\caption{The maximum jump in the Susceptibility $\chi$ for each year ($1979$-$2018$). El Ni\~{n}o years are shown in  red with a  gradation according to their strength. There have been $12$ such El Ni\~{n}o episodes out of which ($1982$, $1997$ and $2015$) are very strong El Ni\~{n}o years, ($1987$ and $1991$) are strong El Ni\~{n}o years , ($1986$, $1994$, $2002$ and $2009$) are moderate El Ni\~{n}o years and ($2004$, $2006$ and $2014$) are weak El Ni\~{n}o years.  }
\end{figure}
\par We adopt the following criterion for the prediction of the El Ni\~no. When the maximum jump in the susceptibility crosses a threshold value ($7.1$) an El Ni\~{n}o episode is predicted in the following year. Using this threshold value, nine events are predicted correctly out of twelve with just one false alarm. Almost all the events (very strong, strong and moderate El Ni\~{n}o events) can be predicted by our criterion except for a few weak El Ni\~{n}o events. The threshold has been set to be the value of the jump of the year 2005, which is the indicator year for a weak El Ni\~{n}o year. This minimizes the false alarms. If our results are compared with the normalized largest cluster used  as an indicator (as per Ref.\cite{1}) , it is seen that when $\Delta s_1$ crosses a threshold value ($0.286$), an El Ni\~{n}o episode is predicted in the next year. It predicts eight events correctly out of twelve El Ni\~{n}o events with two false alarms.

The susceptibility pattern  prior to an El Ni\~{n}o episode is further analyzed by looking at the microtransitions signaled by small jumps in the susceptibility occurring before the phase-transition signaled by the largest jump. The locations of these microtransitions, plotted as a function of $C_i$ can be used to predict the critical value of $C_c$ in the given year.  
\begin{figure}[H]
\begin{tabular}{cc}
\includegraphics[width=7.5cm]{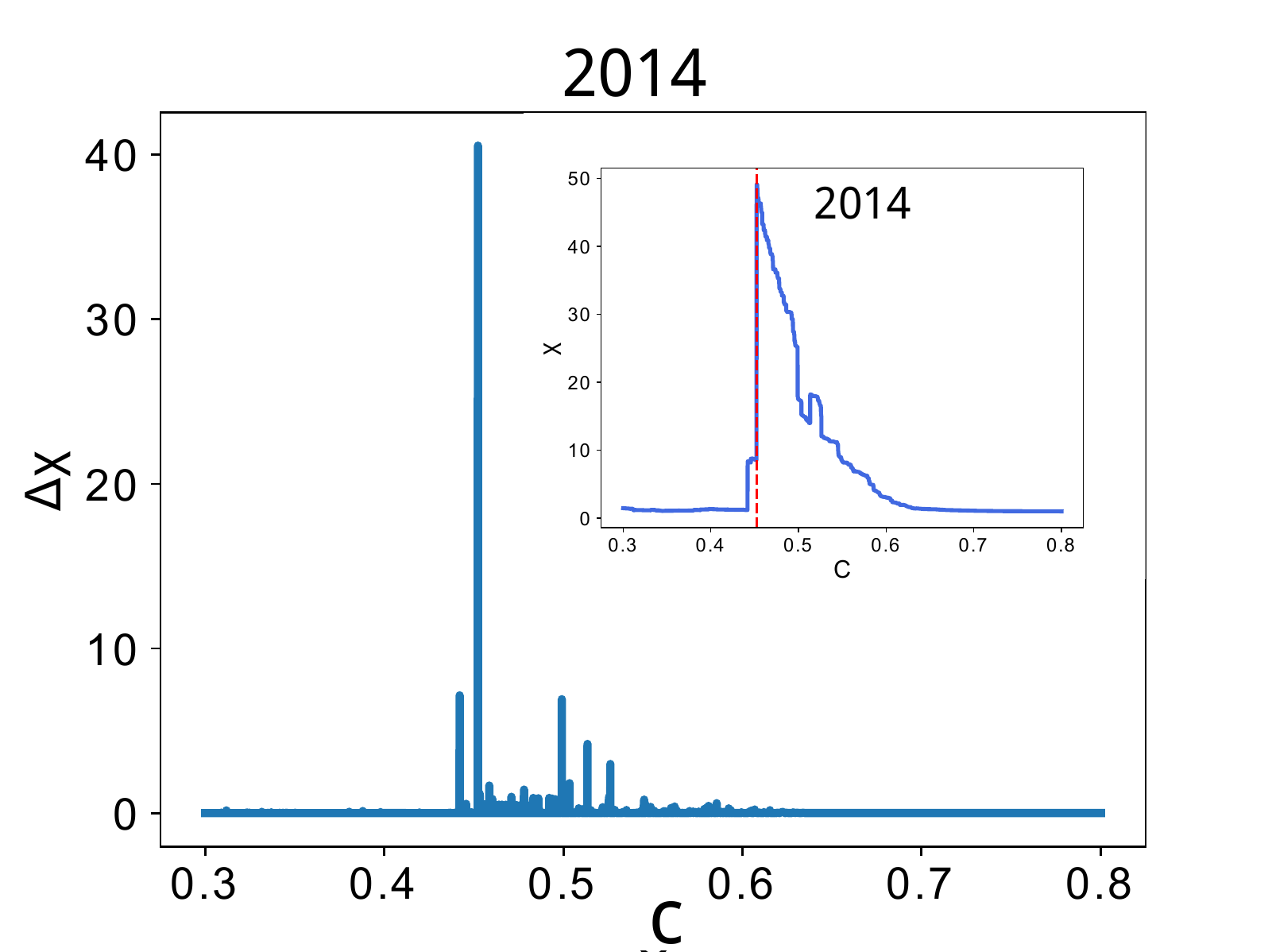} &
\includegraphics[width=7.5cm]{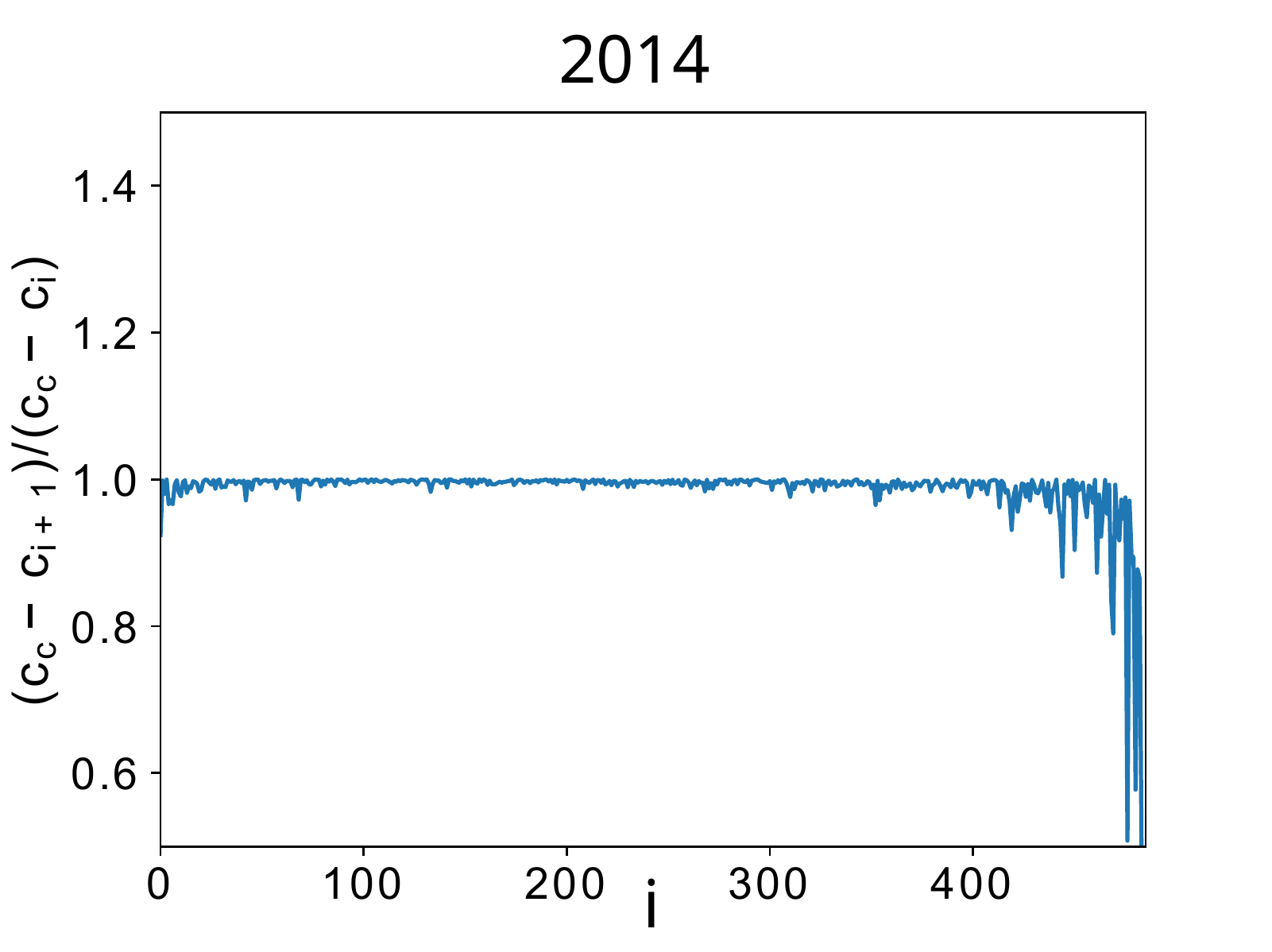} \\
(a) & (b)
\end{tabular}\\
\caption{(a) The jump $\Delta \chi$ as a function of the link strength $C$. (Inset: The susceptibility $\chi$ as a function of the link strength $C$ for the network one year before a very strong El Ni\~{n}o episode, December 2014.) (b) The scaling relation for the relative $\Delta \chi$  positions, $\left(C_{c} - C_{i+1}/C_{c} - C_{i}\right)$ as a function of the index $i$.}\label{sus}
\end{figure}

\par The red line in the susceptibility plot indicates the value of the link strength $C_{c}$ at which the climate network undergoes the percolation transition i.e. the formation of a giant component (cluster) containing $O(N)$ nodes [Fig.\ref{sus}(Inset)]. In [Fig.\ref{sus}(a)] the magnitude of successive jumps in  the susceptibility with respect to the link strength is plotted. Using the scaling relation plotted in [Fig.\ref{sus}(b)] one can write,
\begin{equation}
\frac{C_{c} - C_{i+1}}{C_{c} - C_{i}} = 0.9956
\end{equation}
  This scaling relation enables us to predict the transition point at which the network undergoes a percolation transition. We analyze these microtransitions for each network for each year within the period $1979$-$2018$. It is noted that for the indicator year $2014$, $C_{c} = 0.4669$ using the values  $i = 44, C_{44} = 0.6719$ and $C_{45} = 0.6710$. The value of $C_{c}$ from the scaling relation is in reasonable agreement with the actual value of $C_{c} = 0.4522$ from the order parameter jump. A similar pattern of activity is observed for all the pre-El Ni\~{n}o years. One can predict the transition point using the above scaling relation and starting from $i \approx 45, C_{45} \approx 0.66$ for all the pre-El Ni\~{n}o years in this data. Thus a distinct pattern of microtransitions is observed before the percolation phase transition point for all the indicator years prior to an El Ni\~{n}o activity. We notice some crucial points regarding these microtransitions in the indicator years such as, the $C_c$ value ranges from  $0.4411$ to $0.4736$  for all indicator years except for the indicator year $2008$ prior to moderate El Ni\~{n}o episode. Again, $| \Delta \chi |$ the jump in the susceptibility is very high ranging between $40$ to $100$ for indicator years compared to the other years. 
The  topological analysis of all the $39$ climate networks has also been carried out and the histogram of the link lengths $\ell_{i,j}$ (i.e. the geographical distance between point $i$ and point $j$) observed in the network is plotted. These remote connections are typically of the order of thousands of kilometers and have been called  teleconnections \cite{prl5}. Teleconnections are of great importance in climate dynamics as they reflect the underlying climatic currents mirroring the transportation of energy. These teleconnections between various remote regions have different weights associated with them. It can be seen that the number of large distance links increase substantially during El Ni\~{n}o and La Ni\~{n}a activity. 

\begin{figure}[H]
\begin{tabular}{c c}
\setlength{\tabcolsep}{-60pt} 
\renewcommand{\arraystretch}{0.0} 
\includegraphics[width=0.5\textwidth]{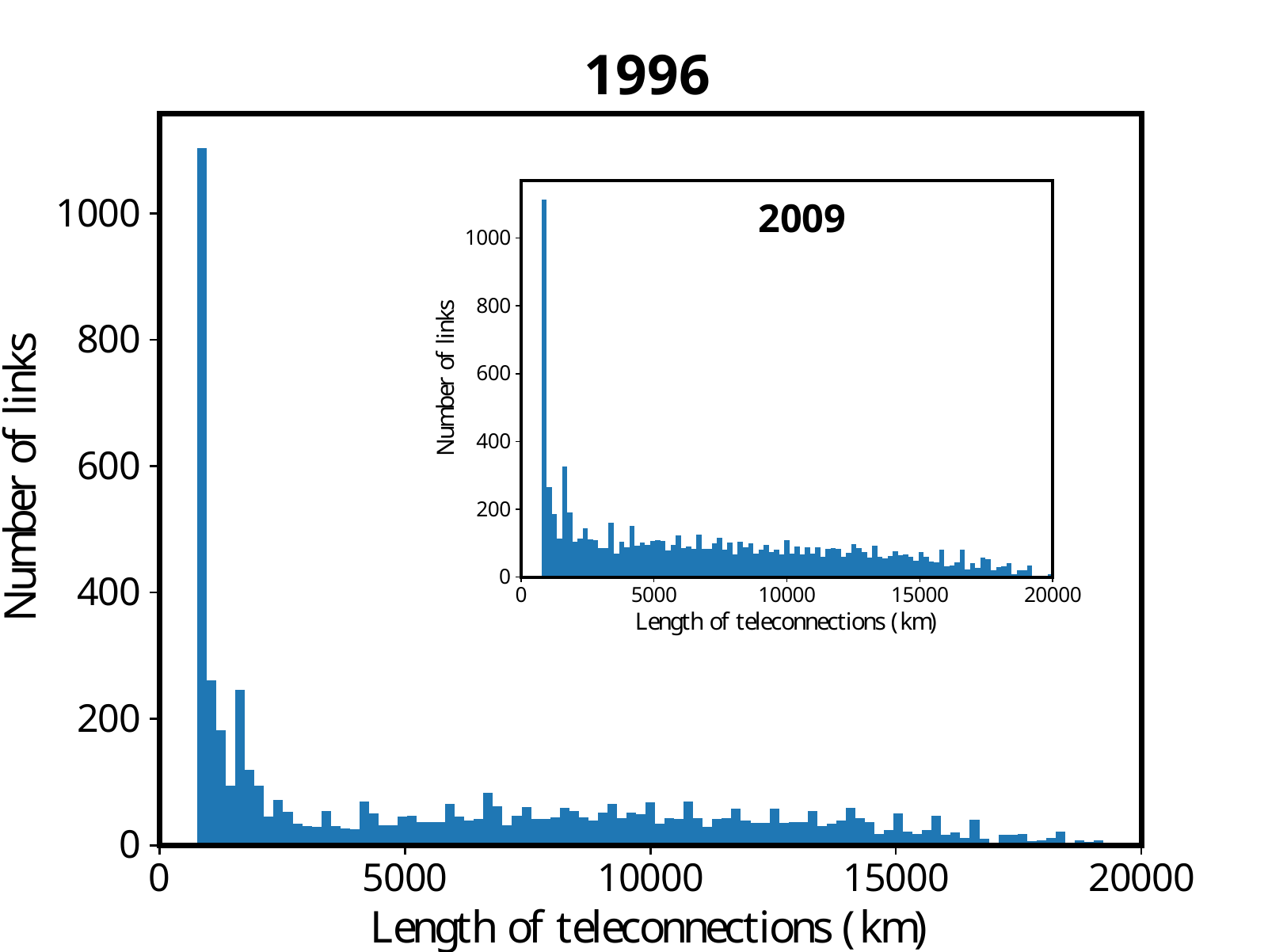}  &
\includegraphics[width=0.5\textwidth]{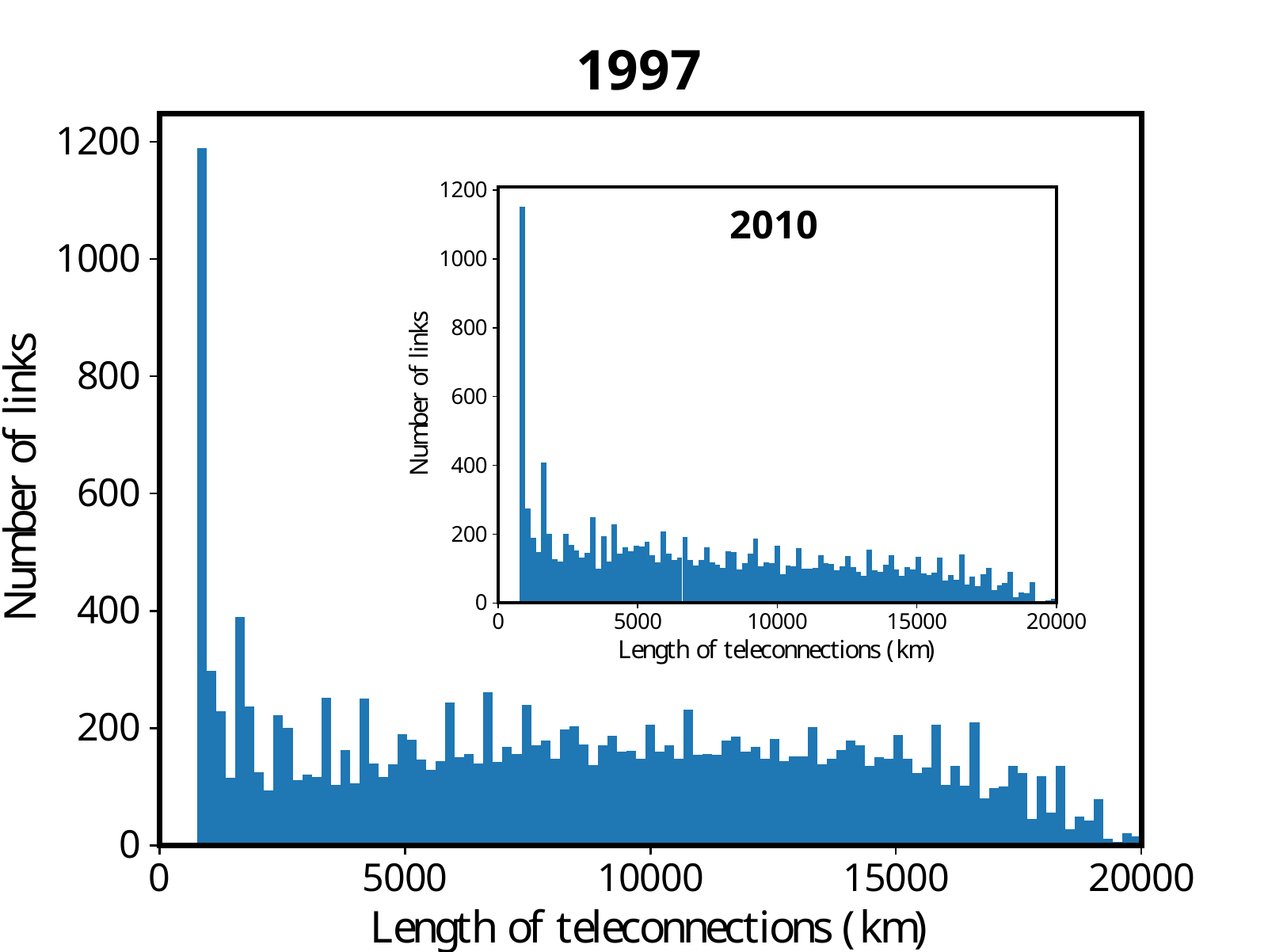}   \\
(a) & (b) \\
\end{tabular}
\caption{The histogram of the teleconnections in kilometers for the network (a) One year before a very strong El Ni\~{n}o episode, December 1996. (Inset: One year before strong La Ni\~{n}a episode, December 2009). (b) During a very strong El Ni\~{n}o episode, December 1997. (Inset: During a strong La Ni\~{n}a episode, December 2010).}
\end{figure}  

\par
\begin{figure}[H]
\setlength{\tabcolsep}{-13pt} 
\renewcommand{\arraystretch}{0.0} 
\begin{tabular}{ c c c c}
\includegraphics[height = 4.8cm]{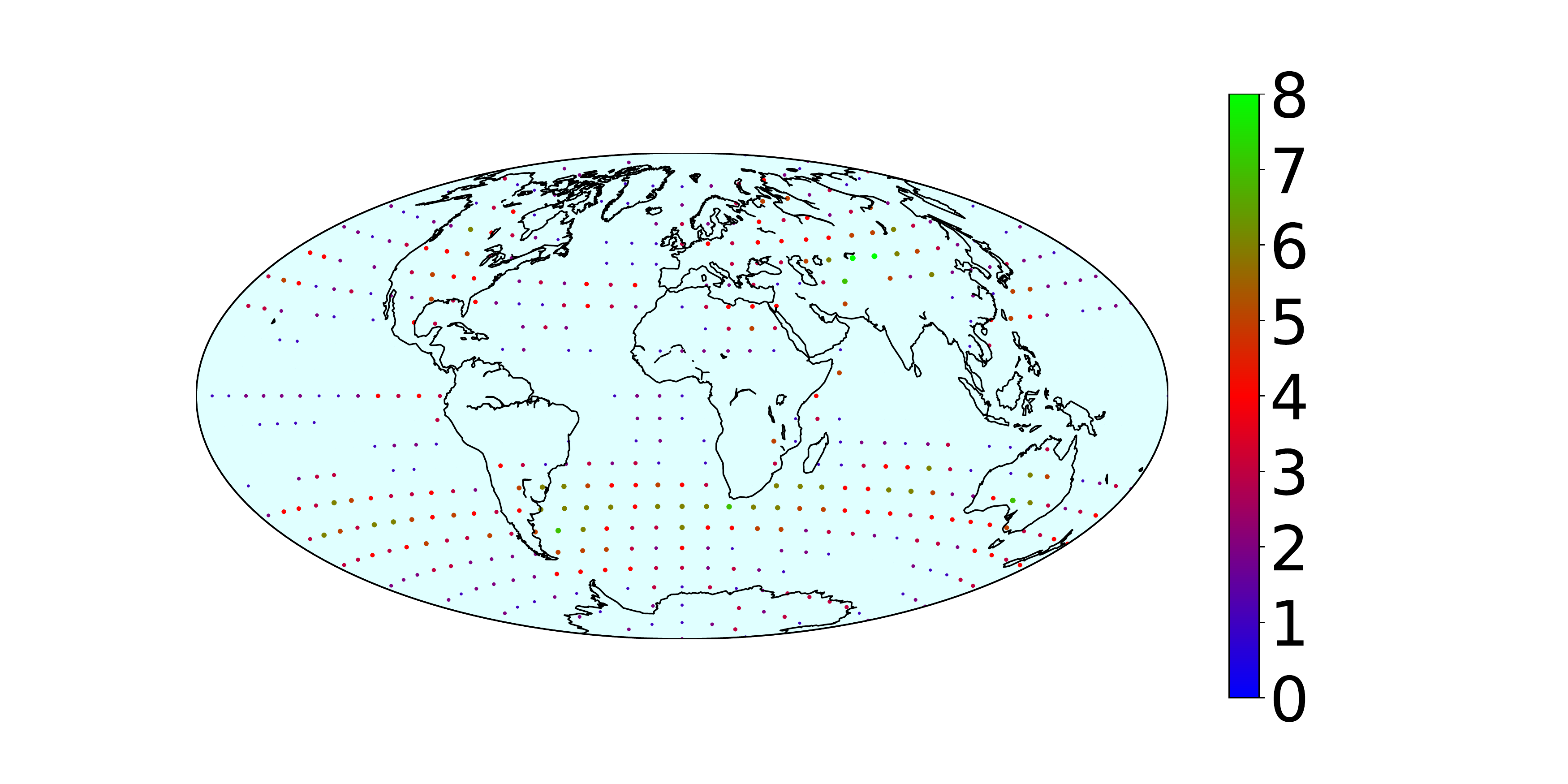} & 
\includegraphics[height = 4.8cm]{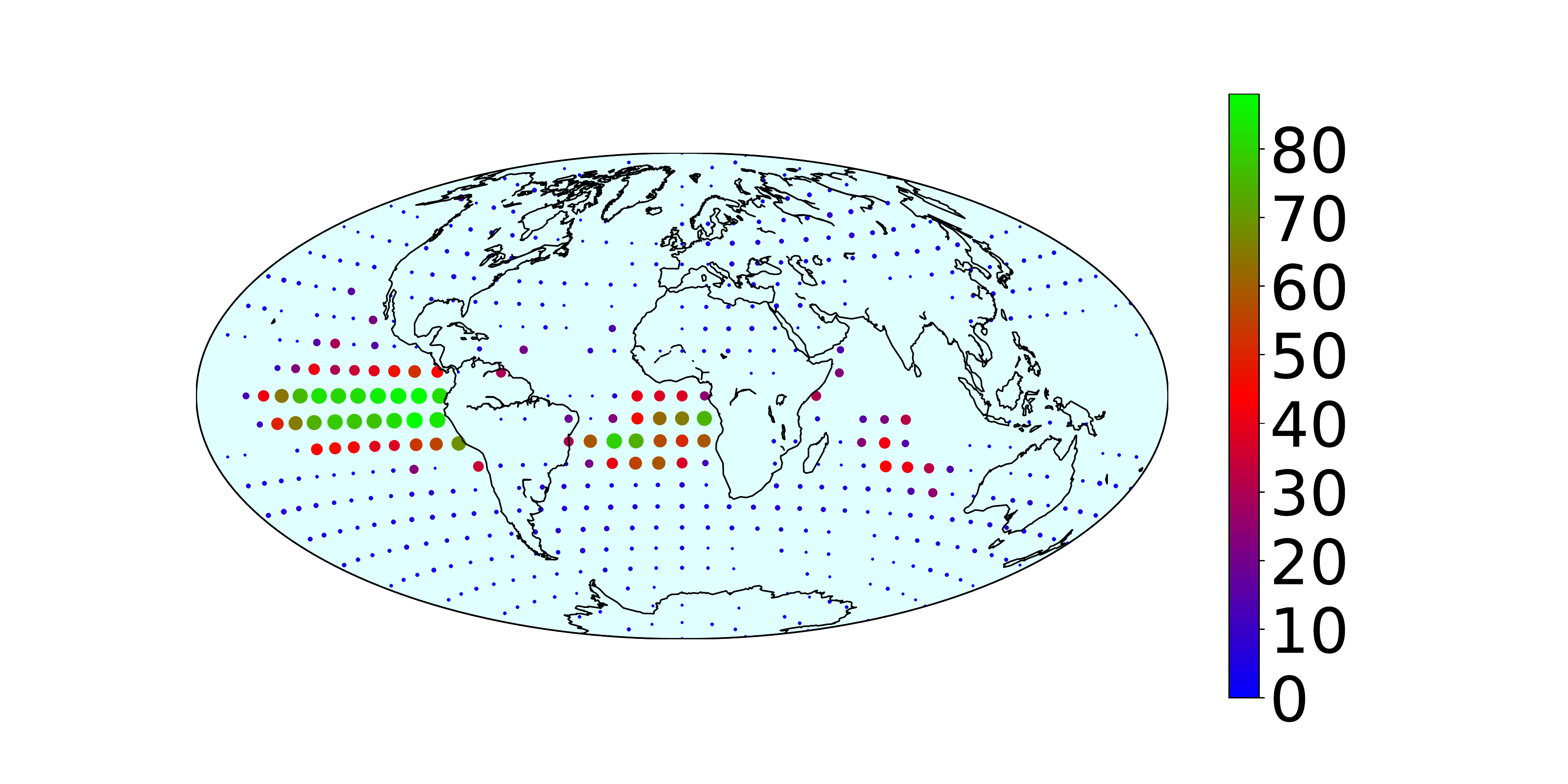} &  \\
(a) & (b) \\
\includegraphics[height = 4.5cm]{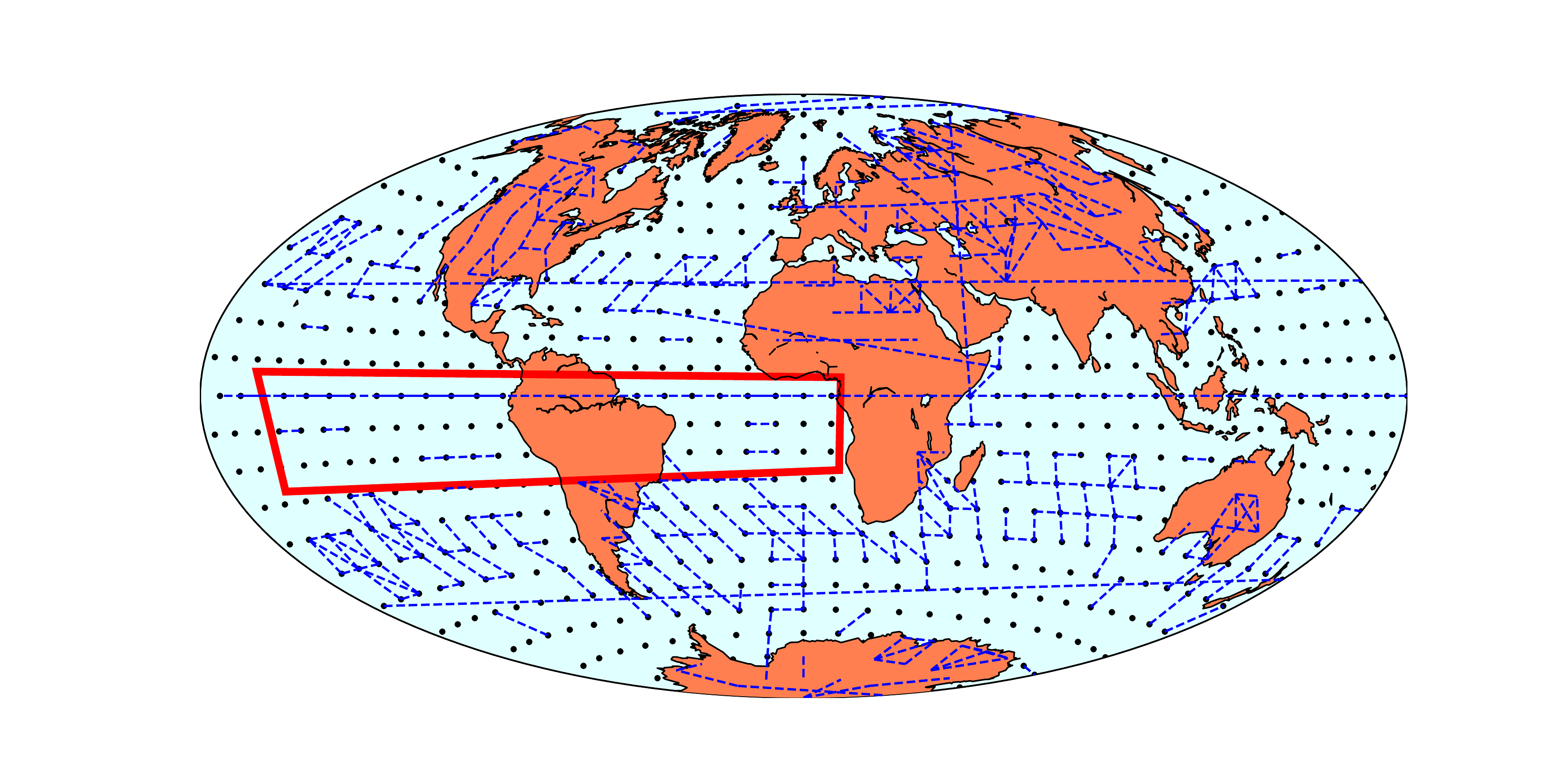} & 
\includegraphics[width = 8.5cm, height = 3.5cm]{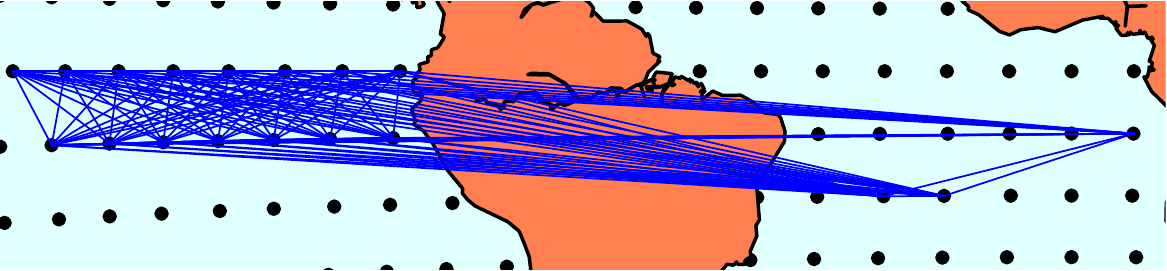} &  \\
(c) & (d) \\
\end{tabular}
\caption{(i) The geographic distribution of degrees for the nodes of the network for  $C = 0.5$  and above (a) One year before a very strong El Ni\~{n}o episode, December 1996. (b) During very strong El Ni\~{n}o episode, December 1997. (ii) The Teleconnection plots of the networks at $C = 0.5$ and above (c) One year before a very strong El Ni\~{n}o episode, December 1996. The same quantity in the  red rectangular region in (c) is shown during a very strong El Ni\~{n}o episode, December 1997 in (d) indicating  the teleconnections between the nodes with degree $\geq 70$.} \label{tel}
\end{figure}

 The above figures indicate that the total number of teleconnections increase significantly during El Ni\~{n}o as well as La Ni\~{n}a activity. An increased number of teleconnections with higher value of correlation strength are also seen for El Ni\~{n}o years. The degrees of nodes (number of links per node) at each grid location has been plotted in Figure [\ref{tel}(c)] to analyze the geographical distribution of nodes of high degree. 

It can be clearly observed from above plots that during very strong El Ni\~{n}o activity, nodes from the El Ni\~{n}o basin have higher degree ($\geq 40$) and are strongly connected whereas for indicator year nodes have a lower degree distribution range ($0$-$15$) and highly connected nodes appear outside the El Ni\~{n}o basin. We thus have a set of topological signatures for the indicator years of the El Ni\~no episodes in addition to the susceptibility and the microtransitions. 
\begin{table}[H]
\begin{tabular}{|l|l|l|l|l|l|l|l|l|l|}
\hline
 & Year       & $C_c$                     & $C_{max}$   & $\Delta\chi $ & Total \\ & & & & & number \\
& & & & & of links \\
\hline
\bf{1.} & \bf{1981}   & 0.4411                 & 0.7609 & 22                         & 10108              \\
& 04/1982 to 06/1983, (15 months)  & 0.5419        & 0.8863 & 3                       & 24850              \\

\bf{2.} & \bf{1985}      & 0.4536                 & 0.796  & 30                         & 10601              \\
& 09/1986 to 02/1988, (18 months)                   & 0.4879 & 0.9361 &  5                   & 20021              \\
 
\bf{3.} & \bf{1990}            & 0.4682                 & 0.7912 & 18                         & 10592              \\
& 05/1991 to 06/1992, (14 months)                   & 0.4874         & 0.8513 & 6                     & 16545              \\

\bf{4.} & \bf{1993}            & 0.4695                 & 0.7787 & 15                         & 11981              \\
& 09/1994 to 03/1995, (7 months)                    & 0.4871         & 0.8538 & 4                       & 14099              \\
 
\bf{5.} & \bf{1996}        & 0.4736                 & 0.843  & 37                         & 10331              \\
& 05/1997 to 05/1998, (13 months)                         & 0.5027         & 0.952  & 5                       & 27469              \\

\bf{6.} & \bf{2001}         & 0.466                  & 0.8157 & 3                          & 10122              \\
& 06/2002 to 02/2003, (9 months)                    & 0.4856         & 0.7986 & 5                      & 14311              \\

\bf{7.} & \bf{2003}             & 0.4856                 & 0.7984 & 5                          & 9393               \\
& 07/2004 to 02/2005, (8 months)                   & 0.4660         & 0.7929 & 7                       & 10000              \\

\bf{8.} & \bf{2005}              & 0.466                  & 0.7929 & 7                          & 10000              \\
& 09/2006 to 01/2007, (5 months)                   & 0.4948         & 0.8712 &  2                       & 12830              \\

\bf{9.} & \bf{2008}          & 0.5126                 & 0.8346 & 6                          & 17882              \\
& 07/2009 to 03/2010, (9 months)                   &  0.5119         & 0.8747 & 1                       & 24397              \\

\bf{10.} & \bf{2013}   & 0.462                  & 0.7877 & 27                         & 11170              \\
& 11/2014 to 05/2016, (19 months)                   & 0.5022 & 0.8949 & 3                    & 20301              \\

\bf{11.} & 2018                    & 0.5001                 & 0.865  & 4                          & 12919              \\
\hline
\end{tabular}
\caption{All El Ni\~{n}o events between $1979-2018$ with distinct precursors.}
\label{ind}
\end{table}
\par
Table [\ref{ind}] summarizes all the distinct and significant indicators for all the El Ni\~{n}o episodes occurring  between  $1979$ to $2018$. Indicator years are shown in bold font followed by the corresponding El Ni\~{n}o episode or event with its duration. As can be seen, lengths of the episodes vary from a minimum of $5$ months to $19$ months. It is observed that all the indicator years have significantly higher $| \Delta \chi |$ values. We see that when the quantity $| \Delta \chi |$ crosses a threshold value $7$, it can be used as an alarm and it is predicted that an El Ni\~{n}o event will start in the following calendar year. A network is constructed for each year using the data of $\approx 565$ days and  a prediction is made at the end of every year ($31^{st}$ December). The lead time between the prediction and the beginning of the El Ni\~{n}o episodes is $\approx 6.42$ months and $\approx 9.57$ months till the time when the episode peaks. It is also observed that magnitude of $| \Delta \chi |$ is distinctly higher for the indicator years followed by strong and very strong El Ni\~{n}o episodes compared to weak or moderate episodes. The value $C_c$ at which the network undergoes a discontinuous percolation transition is given in [table \ref{ind}]. Here, the difference between indicator years and El Ni\~{n}o years can be clearly seen. For indicator years the $C_c$ value generally ranges between $0.4411$ to $0.4736$ whereas during the El Ni\~{n}o years $C_c$ is shifted towards higher values ($ > 0.5$). The total number of links for all the values of $C$ in indicator years is about $10,000$, whereas in El Ni\~no years it is between $14,000-20,000$. We observed that when the total number of links crosses the threshold ($8100 - 9500$), one can expect El Ni\~{n}o in the following calendar year. The number of highly correlated links, with $C$ values $C_{max}$, up to $0.9$ is seen in networks  of El Ni\~{n}o years. The node degrees in these years also take high values, as can be seen in Fig.[\ref{tel}(b)]. Some of these quantities can be used as precursors in conjunction with the susceptibility jumps and $C_c$ values in years where the El Ni\~no phenomena is mild.  
Thus, we have identified a comprehensive set of network quantities which can act as clear precursors of El Ni\~no years, in climate networks constructed using surface air temperature data 
from $1979$ to $2018$. We note that networks constructed for other years do not possess these distinctive features. The El Ni\~no prediction can be made between 4 to 9 months in advance. We also note that our precursors (the jump in the susceptibility $\Delta \chi$ and critical correlation value) miss the prediction of an upcoming moderate El Ni\~no event in the year $2009$-$10$. Nevertheless we observe that out of all the indicators, two indicators i.e. the value of $C_{max} =0.8346$  and the total number of links$= 17882$ do suggest an upcoming El Ni\~no event. The node degree plot of the year $2008$ (not shown here), shows nodes of degrees as high as $40$, which also suggests some indication of the $2009$-$10$ El Ni\~no event. Only one false positive alarm is observed in the year $2000$ in the $39$ climate networks studied, where the susceptibility and $C_c$ predicted an El Ni\~no in the year $2001$ which did not actually occur. 
This was due to the presence of a prolonged and strong La Ni\~na episode
which lasted $32$ months between July $1998$ to February $2001$. The total number of links and the highest degrees of nodes showed moderate values
viz. $11000$ and $16$ in the year $2000$ and were not indicative of an El Ni\~no phenomenon in $2001$. We also observed from the table [\ref{ind}] that year $2018$ has slightly higher values of precursors such as $C_{max}$ ($0.865$), total number of links ($12919$) and nodes of degrees as high as $30$. These values of precursors indicated a weak El Ni\~no episode which was observed in year $2019$.
It may be possible to remove these anomalies by using a sliding window for the time averages, or by sampling focused geographic regions. These may also help to identify the tipping points and tipping locations of the phenomena. We also observed stronger link density in the indicator years in the  area of polar jet streams, i.e. powerful meandering air currents that arise due to atmospheric heating by solar radiation and the action of the Coriolis force in a generally westerly direction.
It appears that jet streams get stronger in the indicator years and show a different pattern for these years depending on the beginning of El Ni\~{n}o (Spring or fall) \cite{elena}. We hope to quantify and provide further support for this assertion in future work.

To summarize we have identified  a set of precursors for the El Ni\~no phenomena, using the construction of climate networks. These include the susceptibility, the value of the critical correlation, the maximum value of the correlation strength and total number of links observed in the climate network. The topological characterizers i.e. the degree distribution and the total number of links, supplement the information available in the order parameter and the susceptibility. These precursors, taken together, constitute signatures of the indicator year which can  reliably predict an El Ni\~no event, four to ten months in advance, and significantly reduce the frequency of false alarms. We hope our methods provide pointers for other investigations in the context of climate networks.
\section{Acknowledgement}
 RS thanks IIT Madras for an Institute postdoctoral fellowship for the period in which this work was done.


\end{document}